# Investigation of Non-Radiative Relaxation Dynamics Under Pulsed Excitation Using Photon Absorption Remote Sensing: A Proof-of-Principle Study in Mechanical Sensing

Channprit Kaur [a], Aria Hajiahmadi[a], Benjamin R. Ecclestone, James E.D. Tweel, James A. Tummon Simmons, Parsin Haji Reza*

[a]these authors contributed equally.

*Photomedicine Labs, Department of System Design Engineering, University of Waterloo, Ontario, Canada*
*Corresponding author: phajireza@uwaterloo.ca

## 1. Abstract

The mechanical properties of micro-scale bio-entities are fundamental for understanding their functions and pathological states. However, current methods for assessing elastic properties at single-particle level such as Brillouin and atomic force microscopies exhibit intrinsic limitations, including being often slow, having poor resolution, or involving complicated and invasive setups. In this study, we explore Photon Absorption Remote Sensing (PARS) microscopy as a unique solution for mechanical sensing of single micro-objects. PARS uses probe beam scattering/reflectivity measurements to capture non-radiative relaxation process following the absorption of a pulse of light by a micro-object. In particular, we demonstrate that, when operating at GHz-range bandwidth, PARS can trace the sub-nanosecond dynamics of non-radiative relaxation in individual micro-objects, capturing both photoacoustic (PA) pressure propagation and thermal diffusion. This GHz-range measurement, in conjunction with a developed descriptive model, enables the experimental extraction of a minimally distorted PA temporal profile. The PA temporal profile contain information on the ratio between the absorbing object's sound speed and its characteristic diameter, offering a new dimension in PARS microscopy. This enables the assessment of the object's elastic properties, deduced from its speed of sound. Additionally, it offers the potential for sizing objects with known sound speeds. The proof of principle experiments was conducted using spherical polystyrene absorbers, ranging in size from 1 to 10 micrometers with known properties, embedded in a Polydimethylsiloxane (PDMS) matrix. This technique expands the scope of PARS imaging, opening new perspectives for clinical applications in mechanobiology by demonstrating its potential for mechanical imaging.

## 2. Introduction

Understanding the mechanical properties of individual micro-bio-entities, such as biological cells, organelles, and micro-organisms holds significant importance in cancer research, disease diagnosis and tissue engineering [1-5]. However, probing the elastic properties of micron-scale bio-entities at the single-particle level within a heterogeneous medium, such as biological tissues, poses significant challenges. This process requires molecular-specific mapping to accurately locate the target micro-entities amidst the surrounding biological matrix [6]. To address this, a combination of molecular- and mechanical-specific imaging is required for simultaneous mechanochemical mapping of samples. Among the various available approaches, the integration of Raman and Brillouin scattering microscopies has emerged as a promising solution [6-8]. These methods provide appropriate spatial resolution for assessing single micro-objects while being non-contact, non-invasive, and label-free. However, both Raman and Brillouin scattering suffer from relatively low scattering cross-sections even in their stimulated version [9, 10]. This limitation restricts their application in biological imaging [11].

Alternatively, absorption cross-sections of infrared and ultraviolet-visible spectra are orders of magnitude larger than those of Raman and Brillouin scattering [9, 12]. In optical absorption, the absorbed incident electromagnetic energy relaxes through the combination of radiative and non-radiative mechanisms [13]. These mechanisms underscore the potential of optical absorption-based techniques for both molecular and mechanical contrast imaging. In the radiative relaxation process, the fluorescent emission contrast is directly detected from Stokes-shifted emission with diffraction-limited resolution in autofluorescence microscopies. On the other hand, non-radiative contrast imaging can be realized with high sensitivity indirectly through sensing the resultant local temperature rise or pressure changes in photothermal and photoacoustic microscopies, respectively. However, despite its potential to provide insights into the physical properties of the absorber and its microenvironment, optical absorption-based techniques have not been explored directly and quantitatively for mechanical-specific imaging [14-18]. In particular, the temporal profile of PA pressure theoretically encodes information on sound speed and the geometrical properties of the absorbing object, yet this potential remains largely unexplored.

In photoacoustic microscopy (PAM), the pressure generated by a pulse of excitation light is detected using an ultrasound transducer. However, the finite bandwidth of ultrasound transducers (typically 50% of the central frequency) distorts the measured temporal profile of the PA wavelet at each spatial location. This distortion complicates the interpretation of the mechanical and geometrical properties of the absorbing object from the measured PA signal in the time domain [19]. To address these limitations, several optical ultrasound sensors, including optical micro-ring resonators [20], Fabry–Perot interferometers [21], and surface plasmon resonators[22], have been developed in recent years. These sensors offer broad bandwidths that help mitigate the said limitations in PA sensing. Nonetheless, these sensors are typically positioned at a certain distance from the absorption site, leading to suppression of high-frequency components and again distortion of the PA wavelet [23]. Thus, to accurately capture the PA profile in the time domain, both broadband detection and close proximity of the sensor to the absorbing object are essential.

Photothermal microscopes offer an alternative approach to gauge the non-radiative relaxation process. In these microscopes, pump-probe configuration is used, in which a probe beam outside the absorption band is co-focused with the excitation spot. After excitation, the probe may undergo divergence/convergence, intensity modulation, phase delay modulation, or a combination of these effects as the result of an increase in the temperature of the absorber and its microenvironment [24-26]. By leveraging these probe beam modulations, the pump-probe configuration enables the characterization of non-radiative relaxation —primarily as temperature rise— directly at the absorbing object's site without any additional separation. However, these methods often fail to generate adequate photoacoustic (PA) pressure due to insufficient excitation speed. To overcome this limitation, photoacoustic remote sensing was introduced in 2017 by Haji Reza et al [27]. This technique selected the excitation sources based on the photoacoustic confinement time, allowing the simultaneous generation of both pressure and temperature modulations.

Combining the strengths of radiative (e.g., autofluorescence), and non-radiative (e.g., photothermal, and photoacoustic) techniques into a single modality, our group recently introduced photon absorption remote sensing (PARS) which has evolved from a photoacoustic remote sensing method [28-30]. PARS employs a short-pulsed excitation laser and a confocal probe to capture all non-radiative induced optical modulations (photothermal and photoacoustic), while radiative emissions are measured directly by capturing Stokes shifted photons. These contrasts have demonstrated exceptional molecular specificity across various samples [31, 32]; however, the potential for deriving mechanical contrast remains unexplored.

In PARS, PA pressure-induced modulations can be directly observed with the confocal probe, eliminating distortion caused by frequency-dependent attenuation during wave propagation —a common limitation in traditional photoacoustic methods. However, earlier PARS imaging systems utilized detection bandwidths in the MHz range, restricting their ability to fully capture PA pressure dynamics. This bandwidth limitation introduced significant time-domain distortion in the signal response to the rapid variations of pressure propagation dynamics.

In this work, we address this limitation by upgrading the detection bandwidth of the non-radiative PARS (NR-PARS) submodule to the gigahertz (GHz) range. This enhancement enables capture of the full dynamics of non-radiative relaxation in real time, including both pressure propagation and heat diffusion. In order to extract the PA pressure response in the time domain, a model is developed to describe how pulsed excitation-induced PA pressure, along with thermal effects, generates the NR-PARS signal. This descriptive model enables isolation of the PA-induced signal contribution, which can then be used to extract the ratio between the absorber's sound speed and its characteristic diameter. A proof-of-principle experiment was conducted using polystyrene spheres of varying sizes (1 µm, 3 µm, 6 µm, and 10 µm) embedded in a Polydimethylsiloxane (PDMS) matrix. The proposed method demonstrates strong potential for biomechanical imaging at the single-particle level, adding a mechanical property contrast to PARS microscopy. This complements the existing molecular-specific imaging capabilities of PARS microscopy, enhancing its utility for biological applications.

### 3. Theoretical model for epi-detected NR-PARS signal in idealized samples

To explore signal mechanism of NR-PARS submodule in a reflection mode design (see Fig. 1(a)), we consider an absorbing object (a micro-sphere) embedded in an infinite, homogeneous, non-absorbing medium as an idealized sample. It is also assumed that upon excitation by an ultrashort pulsed laser, the micro-object undergoes instantaneous and uniform heating. This instantaneous heating triggers non-radiative relaxation response. The resulting temperature and PA pressure alternations in the object and its medium is read out by probe beam's scattering or reflectivity measurements from the absorbing object. In this work, the back-scattered or back-reflected probe light from the object of interest is referred to as the signal field, $\boldsymbol{E_s}(t)=E_s(\text{t})\exp(i\phi_s(t))$, where $E_s(\text{t})$ and $\phi_s(t)$ are the amplitude and phase of $\boldsymbol{E_s}(t)$. The NR-PARS detection scheme exclusively monitors signal field amplitude changes, $E_s(\text{t}) = E_s + \delta E_s(t)$. More specifically, the non-radiative relaxation process modulates existing signal field's amplitude ($E_s$) by an amount $\delta E_s(t)$, which constitutes the NR-PARS signal, $S_{NR-PARS}(t) \propto \delta E_s(t)$ (as depicted in Fig. 1(b)). In our model for the signal, pulsed excitation-induced deflection of the probe beam and lensing effects (e.g., thermal and pressure lensing) are neglected, as they can be captured mostly in transmission mode configurations and in the presence of an effective pinhole only. Therefore, we assume that the modulation of the signal field is governed only by perturbation of refractive indexes of the object ($\delta n_p$) and medium ($\delta n_m$) as well as object's volume ($\delta V_p$), where $n_p$ and $n_m$ are equilibrium refractive indices of absorbing object and medium, respectively, and $V_p$ is the volume of the object before excitation incident (see Fig. 1(b)). To theoretically derive the time-resolved NR-PARS signal, the following two steps are performed:

**Step (i)**- Solving spatiotemporal evolution of temperature, pressure, and resulting strain fields: We compute the PA pressure and temperature rise fields generated by the non-radiative relaxation process. From pressure and temperature fields, we derive the resulting strain field, which is defined as $\eta(\mathbf{r}, t) = \nabla . \xi(\mathbf{r}, t)$ where ξ(**r**, *t*) represents the displacement field.

**Step (ii)**- Relating the strain field to signal field's amplitude modulations: we relate the strain field $\eta(\mathbf{r}, t)$ to changes in the object's volume ($\delta V_p$) and the perturbation in refractive indices ($\delta n_p$ and $\delta n_m$). Finally, within the framework of the relevant scattering/reflection theory, we obtain the time-resolved variations of signal field's amplitude ($\delta E_s(t)$) based on $\delta n_p$, $\delta n_m$ and $\delta V_p$.

These steps will be discussed in detail in the subsequent sections.

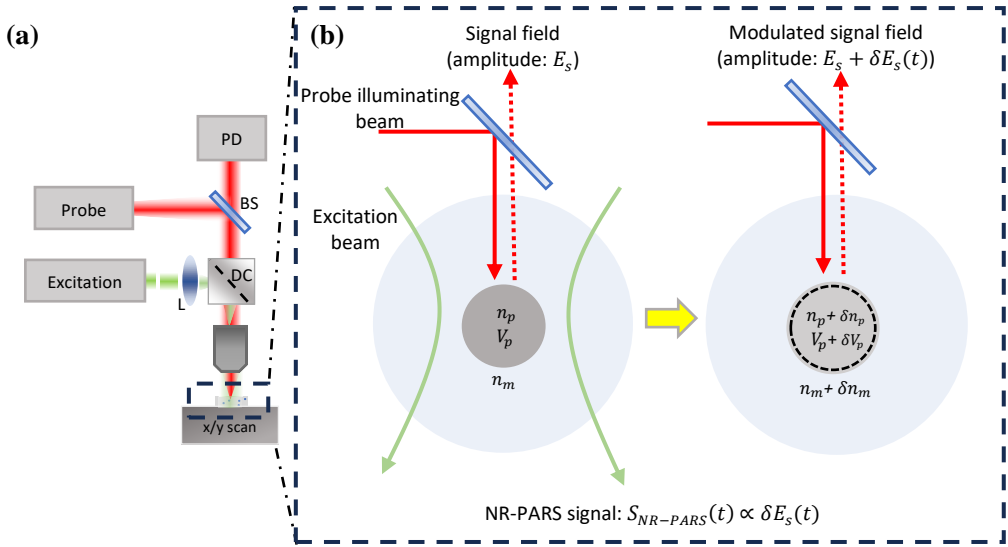

**FIG. 1: Principle of epi-detected NR-PARS submodule and proposed theoretical model for its signal. (a) Simplified optical system of NR-PARS submodule in reflection mode design: BS (Beam Splitter), DC (Dichroic Mirror), PD (Photodiode), L (lens), OL (Objective Lens), S (Sample), x/y scan (translation stage) (b) NR-PARS signal mechanism for an absorbing particle embedded in a homogeneous non-absorbing medium, based on modulation of the amplitude of signal field (i.e., backscattered or back-reflected probe beam). The modulation may arise from perturbation of refractive indices of absorber ($\delta n_a$) and medium ($\delta n_m$) and absorber's volume ($\delta V_p$), resulting from the non-radiative relaxation following the absorption of excitation pulse.**

3.1. Step (i): Solving spatiotemporal evolution of temperature, pressure, and resulting strain fields

To solve for the PA pressure and heat generation, along with their spatiotemporal evolution following the absorption of an ultrashort light pulse in our idealized sample, we begin by recalling the basic relevant equations. Neglecting viscosity, the pressure and temperature fields are governed by the following coupled equations, which are applied in both the absorber and medium domains [33]

$$\rho C \frac{\partial T}{\partial t}(\boldsymbol{r},t) - K\nabla T(\boldsymbol{r},t) = S_v(\boldsymbol{r},t) \quad (1a)$$

$$P(\boldsymbol{r},t) - \frac{1}{v}\frac{\partial^2 P}{\partial t^2}(\boldsymbol{r},t) = -\rho\beta\frac{\partial^2 T}{\partial t^2}(\boldsymbol{r},t) \quad (1b)$$

where $P(\boldsymbol{r}, t)$ and $T(\boldsymbol{r}, t)$ are, respectively, the PA pressure and temperature rise field. The relevant physical properties are the mass density $\rho$, the volume expansion coefficient $\beta$, the thermal conductivity $K$, the specific heat capacity at constant pressure or volume $C$ (they are almost equal in solids and fluids), and the speed of sound $v$. Each of these are defined for the absorbing object and the surrounding medium, denoted by 'p' and 'm' indices, respectively. Also, $S_v(\boldsymbol{r},t)$ represent the volumetric density of optical power converted to heat. Eq (1a) and (1b) are standard heat diffusion and PA wave equations, respectively, with their corresponding source terms on the right side of the equations. The strain field is related to PA pressure and temperature rise field as [34]

$$\eta(\boldsymbol{r},t) = -\kappa P(\boldsymbol{r},t) + \beta T(\boldsymbol{r},t), \quad (2)$$

To obtain the time-resolved strain field, given the source terms of Eq. (1a) and (1b), one must solve the thermal diffusion problem and PA pressure problem in a coupled way, which pose complications and is difficult to interpret. Fortunately, under ultrashort pulsed excitation, heat deposition is fast relative to both PA pressure propagation and heat diffusion. Hence, the strain field within the absorber remain neglectable during the pulse duration. Then, according to Eq. (2), $\eta(\boldsymbol{r}, t = 0) = 0$, we take initial temperature rise ($T_o$) and pressure rise ($P_o = \frac{\beta T_o}{\kappa}$) as initial values for $T(\boldsymbol{r}, t = 0)$ and $P(\boldsymbol{r}, t = 0)$ across the absorber instead of source terms in Eq. (1a) and (1b), respectively. This facilitates separation of Eq. (1b) from the heat diffusion equation (Eq. (1a)). Eq. (1a) and (1b) are then reduced to heat diffusion and propagation of the generated heat and pressure, respectively, as

$$\rho C_p \frac{\partial T}{\partial t}(\boldsymbol{r},t) - K\nabla T(\boldsymbol{r},t) = 0, \quad (3a)$$

$$\frac{1}{v_s^2}\frac{\partial^2 P}{\partial t^2}(\boldsymbol{r},t) - \nabla^2 P(\boldsymbol{r},t) = 0. \quad (3b)$$

Now, the PA pressure $P(\boldsymbol{r},t)$ and temperature rise field $T(\boldsymbol{r},t)$ can be solved independently for a spherical absorber with the diameter $D_p$ in an infinite homogeneous medium. Temperature rise field is solved by numerical simulation in accordance with Eq. (3a) with respect to the thermal properties of both the absorber and the medium (see the Supplementary Note 1 for details of the finite element method (FEM) simulation). Also, according to Eq. (3b), the spatiotemporal profile of pressure is solved analytically. In these solutions, the host medium is assumed to be an ideal medium for both heat and pressure transfer, exhibiting perfect thermal contact and minimal acoustic impedance mismatch with the absorbing object, mirroring the conditions found in biological samples [35].

The resulting PA pressure and temperature rise fields ($P(\mathbf{r},t)$ and $T(\mathbf{r},t)$) are shown in the sequences of Fig. 2(a). The dynamics of pressure propagation and heat diffusion are characterized temporally by the pressure relaxation time, $\tau_{pa}$, and thermal decay constant time, $\tau_{th}$, respectively. The pressure relaxation time, given by $\tau_{pa} = D_p/v_p$, represents the time required for the generated PA pressure to fully exit the absorbing object, neglecting the possible reflected pressure field from the absorber-medium interface. The thermal decay constant, $\tau_{th}$, refers to the time it takes for the average temperature of the absorber to drop to 1/e of its initial value, from $T_0$ to $\frac{T_0}{e}$. Finally, the strain field is derived by combining the thermal-induced strain and PA pressure-induced strain, which are derived directly from the simulated temperature and pressure fields as $\eta_{pa}(\mathbf{r},t) = -\kappa P(\mathbf{r},t)$ and $\eta_{th}(\mathbf{r},t) = \beta T(\mathbf{r},t)$, respectively. The temporal evolution of the strain field, resulting from temperature rise and pressure variations, as well as their superposition (total strain field), is illustrated in Fig. 1(c) as snapshots at various portions of $\tau_{pa}$ and $\tau_{th}$.

At $t = 0$ immediately after heat deposition by the pulsed excitation, the total strain within the absorbing object is zero and become positive over the time interval of $0 < t < \tau_{pa}$. In this time scale, strain dynamics within the absorber are primarily governed by pressure propagation. In contrast, for $t > \tau_{pa}$, the strain field within the absorber is dictated exclusively by heat diffusion, leading to a gradual decrease in strain magnitude.

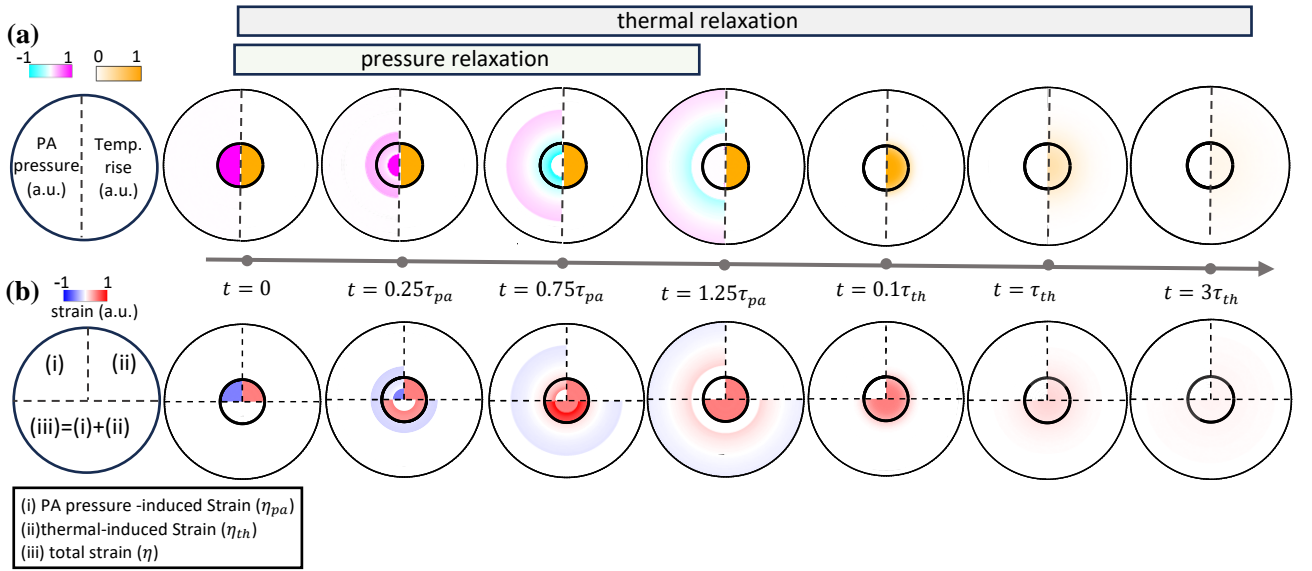

**FIG. 2: Temporal evolution of PA pressure, temperature rise, and resultant strain fields as the result of nonradiative relaxation subsequent to absorption of an ultrashort pulse of optical excitation. (a) shows the snapshot of the temperature rise and pressure fields. (b) shows the snapshots of PA pressure-induced strain, thermal-induced strain as well as their superposition, that is total strain field. The time**

$t = 0$, **first panels in both (a) and (b), corresponds to the time just after heat deposition process of pulsed excitation. The resulting absorber's dimension changes are not illustrated in snapshot panels.**

3.2. Step 2: Relating the strain field to signal field's amplitude modulations

As mentioned in the description of the model, the change in signal field amplitude $E_s(t)$ may arise from perturbation in the refractive indices of absorber and the surrounding medium at their boundary — denoted as $\delta n_p$, and $\delta n_m$ respectively—as well as the volume changes in the absorber, $\delta V_p$. Hence, the general relation for $\delta E_s(t)$ can be expressed as

$$\delta E_s = \frac{\partial E_s}{\partial n_p}\delta n_p + \frac{\partial E_s}{\partial n_m}\delta n_m + \frac{\partial E_s}{\partial V_p}\delta V_p, \quad (4)$$

Perturbations in the refractive indices primarily depend on local density changes at the object-medium interface and are only secondarily influenced by molar refractivity, which is relatively negligible in biological samples [36]. Therefore, the perturbations in the refractive indices, $\delta n_a$ and $\delta n_m$, can be expressed as

$$\delta n_a \approx \frac{dn_a}{d\eta}\eta(\mathbf{R},t), \quad (5a)$$

$$\delta n_m \approx \frac{dn_m}{d\eta}\eta(\mathbf{R},t), \quad (5b)$$

where $\eta(\mathbf{R},t)$ is the temporal profile of the strain at absorber-medium interface, $\mathbf{R} = \left(\frac{D_p}{2}\right)\mathbf{e}_r$ provided that the spherical coordinate is positioned at the center of the absorbing sphere. Strain-dependent refractive index changes, $\frac{dn_a}{d\eta}$ and $\frac{dn_m}{d\eta}$, can be related to thermo-optic coefficient through volume expansion coefficient as $\frac{dn}{d\eta} \approx \frac{dn}{dT}/\beta$ for both object and medium. Thermo-optic coefficients for several materials can be found in the literature of photothermal spectroscopy/microscopy [36].

Also, the volume change of the object can be expressed in relation to the strain field as

$$\delta V_a = \bar{\eta}(t)V_a, \quad (6)$$

where $\bar{\eta}(t) = \int_\Omega \frac{\eta(r,t)dr}{V_p}$ is introduced as the temporal profile of the average strain across the object's volume ($\Omega$). Using Eq. (4), (5) and (6), $\delta E_s(t)$, which constitutes NR-PARS signal, can be expressed in relation to the strain terms $\eta(\mathbf{R},t)$ and $\bar{\eta}(t)$ as

$$S_{NR-PARS}(t) \propto \delta E_s(t) \approx [f\eta(\mathbf{R},t) + g\bar{\eta}(t)] \times E_s, \quad (7a)$$

where $f$ and $g$ factors are introduced as

$$f \approx \frac{1}{E_s}\left(\frac{\partial E_s}{\partial n_a}\frac{dn_a}{d\eta} + \frac{\partial E_s}{\partial n_m}\frac{dn_m}{d\eta}\right), \quad (7b)$$

$$g \approx \frac{1}{E_s}\left(\frac{\partial E_s}{\partial V_p} V_p\right). \quad (7c)$$

To solve for $\delta E_s(t)$, we begin by obtaining the time-resolved strain terms ($\eta(\mathbf{R},t)$ and $\bar{\eta}(t)$). Then, the analysis will be followed by obtaining the $f$ and $g$ factors with respect to the geometrical configuration of the absorbing object and the probe's focus.

For more facile interpretation strain terms, both $\eta(\mathbf{R},t)$ and $\bar{\eta}(t)$ are regarded as the superpositions of PA-induced and thermal-induced strain fields with respect to Eq. (3) as

$$\eta(\mathbf{R},t) = \eta_{pa}(\mathbf{R},t) + \eta_{th}(\mathbf{R},t), \quad (8a)$$

$$\bar{\eta}(t) = \bar{\eta}_{pa}(t) + \bar{\eta}_{th}(t), \quad (8b)$$

where $\eta_{pa}(\mathbf{R},t) = -\kappa P(\mathbf{R},t)$ and $\eta_{th}(\mathbf{R},t) = \beta T(\mathbf{R},t)$, are temporal profile of PA-induced and thermal-induced strains at the particle-medium interface, respectively. Also $\bar{\eta}_{pa}(t) = -\kappa \int_\Omega \frac{P(r,t)dr}{V_p}$ and $\bar{\eta}_{th}(t) = \beta \int_\Omega \frac{T(r,t)dr}{V_p}$ are temporal profile of PA-induced and thermal-induced volume-averaged strains within the absorbing object, respectively. All four temporal profiles are determined based on the solutions of Eq. (4a) and (4b) obtained from numerical/analytical simulation ($P(\mathbf{r},t)$ and $T(\mathbf{R},t)$ fields [Fig. 2(a)]). The temporal profile of $\eta_{pa}(\mathbf{R},t)$ exhibits a bipolar (N-shaped) pattern, corresponding to the pressure amplitude at the object-medium boundary (Fig. 3(a)). The instantaneous pressure at the boundary immediately following the excitation pulse is $\frac{P_0}{2}$, which corresponds to half of the initial pressure within the object. The time-dependant profile of $\bar{\eta}_{pa}(t)$ is a function of the volume-averaged pressure amplitude within the absorber particle. It initiates with an amplitude of $-\kappa P_0$ (representing the initial pressure-induced strain within the object) and subsequently transitions to a positive amplitude as negative pressure dominates across the object (see Fig. 3(b)).

On the other hand, the temporal profiles of $\eta_{th}(\mathbf{R},t)$ and $\bar{\eta}_{th}(t)$, based on model's numerical simulation, are approximated by exponential decay profiles with maximum initial values of $\beta T(\mathbf{R},0)$ and $\beta T_0$, respectively (see Fig. 3(c) and 3(d), respectively). Here, $T(\mathbf{R},0)$ is the instantaneous temperature rise at the boundary between the object and the medium immediately after the excitation pulse event, given by $T_o\gamma$, where $\gamma = [e_p/(e_p + e_m)]$ is temperature scaling factor and $e_p$ and $e_m$ are the effusivities of the absorbing particle and host medium, respectively [37]. In this expression, $\gamma$ can vary between 0 and 1 depending on the thermal effusivities of the absorber and its surrounding medium. The heat diffusion dynamics simulations for various absorber-medium complexes of different sizes were studied in Supplementary Note 2. The superposition of PA-induced and thermal-induced strain profiles yields the time-resolved strain terms of $\eta(\mathbf{R},t)$ and $\bar{\eta}(t)$, shown in Fig. 3(e) and 3(f), respectively. It is noted from Fig. 3(e) that when the $\gamma < 0.5$, an initial negative strain occurs at the absorber-medium boundary immediately after excitation. Conversely, if $\gamma > 0.5$, the boundary strain $\eta(\mathbf{R},t)$ remains positive. The strain temporal profiles in Figs. 3(a)-3(f) are all normalized with respect to $\eta_0 = \kappa P_0 = \beta T_0$.

Now, with the strain terms, $\eta(\mathbf{R},t)$ and $\bar{\eta}(t)$, known for the object of interest (micro-sphere), the $f$ and $g$ factors must be determined to obtain the ultimate NR-PARS signal [according to Eq. (7a)]. To compute the $f$ and $g$ factors, it is necessary to establish a relationship between the governing parameters of the model ($n_p$, $n_m$, $V_p$) and the signal field's amplitude ($E_s$). The relationship for the signal field can be defined in the paradigm of probe light backscattering or reflection, depending on the geometrical configuration of the absorbing object and the probe's focus. As such, two extreme opposite geometrical configurations are considered, as shown in Fig. 3(g), with respect to the size of the object $D_p$ and the diameter of the focused probe beam ($d_{probe}$):

- Extreme geometrical case (i), $D_p \ll d_{probe}$: To obtain an explicit relationship for the $f$ and $g$ factors under this condition, Rayleigh scattering theory is applied to describe the interaction between the object and the focused probe beam. Rayleigh scattering is consistent with sub-diffraction limit objects ($D_p \ll d_{probe}$) where $D_p \ll \lambda_{probe}$ ($\lambda_{probe}$ is probe wavelength in the medium) [38]. This theory allows for relating the scattered field to the size of the absorber as well as refractive indices of the medium and absorber as (absorption at probe's wavelength is neglectable) [39]

$$E_s = 3\epsilon_m V_p \left(\frac{n_p^2 - n_m^2}{n_p^2 + 2n_m^2}\right) E_i \qquad (9)$$

where $E_i$ is the incident probe field's amplitude, and $\epsilon_m$ is the permittivity of the medium. Then, with respect to Eq. (7b), $f$ factor is derived as $f\left(n_p, n_m, \frac{dn_p}{d\eta}, \frac{dn_m}{d\eta}\right) = \frac{6n_m n_p}{(n_p^2 - n_m^2)(n_p^2 + 2n_m^2)}\left(n_m \frac{dn_p}{d\eta} - n_p \frac{dn_m}{d\eta}\right)$. Similarly, from Eq. (7c), the $g$ factor is obtained as unity.

- Extreme geometrical case (ii), $D_p \sim d_{probe}$: In scenarios where the absorbing particles are of the same order of magnitude as or larger than the probe beam diameter ($D_p \sim d_{probe}$), the resulting secondary field from the interaction of the focused probe beam and the object can be approximated by near-normal reflection. The reflected field as a function of refractive indices is

$$E_s = \left|\frac{n_p - n_m}{n_p + n_m}\right| E_i. \qquad (10)$$

Applying Eq. (10) to Eq. (7b), the expression for the $f$ factor is derived as $f\left(n_p, n_m, \frac{dn_p}{d\eta}, \frac{dn_m}{d\eta}\right) = \frac{2(n_m \frac{dn_p}{d\eta} - n_p \frac{dn_m}{d\eta})}{n_p^2 - n_m^2}$. Similarly, from Eq. (7c), the $g$ factor is found to be $g = 0$, as the probe's reflected field in this case is not influenced by the volume of the particle, meaning that its change does not modulate the reflected light from objects where $D_p \gg d_{probe}$. In case (ii), therefore, the sole contributing term in the NR-PARS signal is the perturbation of the refractive index of both the particle and the surrounding medium.

In the above two extreme cases, the $f$ and $g$ factors are derived explicitly in terms of the governing parameters of our model. However, it is not possible to obtain an explicit expression for configurations that fall between these two extremes, and a numerical approach must be employed to determine these factors. The analysis of these two extreme cases indicates that the $f$ factor can take both positive and negative values, depending on the opto-physical properties of the absorbing object and medium. Conversely, the *g* factor is bounded between 0 and 1, with its upper limit corresponding to extreme geometrical case (i). As the size of the absorber becomes comparable to $d_{probe}$, the effect of volume change on signal diminishes and *g* factor approaches zero [Fig. 3(g)]. The *g* factor remains a non-negative function, as a reduction in the size of the absorbing object cannot increase its scattering cross-section or reflectivity. As such, when *f*>0, two strain terms, $\eta(\mathbf{R},t)$ and $\bar{\eta}(t)$, *predominantly* synergize within the signal, whereas when *f*<0 the two terms *predominantly* counteract each other. The term '*predominantly*' is used to acknowledge the possibility of initial negative strain at the absorber-medium interface when $\gamma < 0.5$ [Fig. 3 (a),(c),(e)].

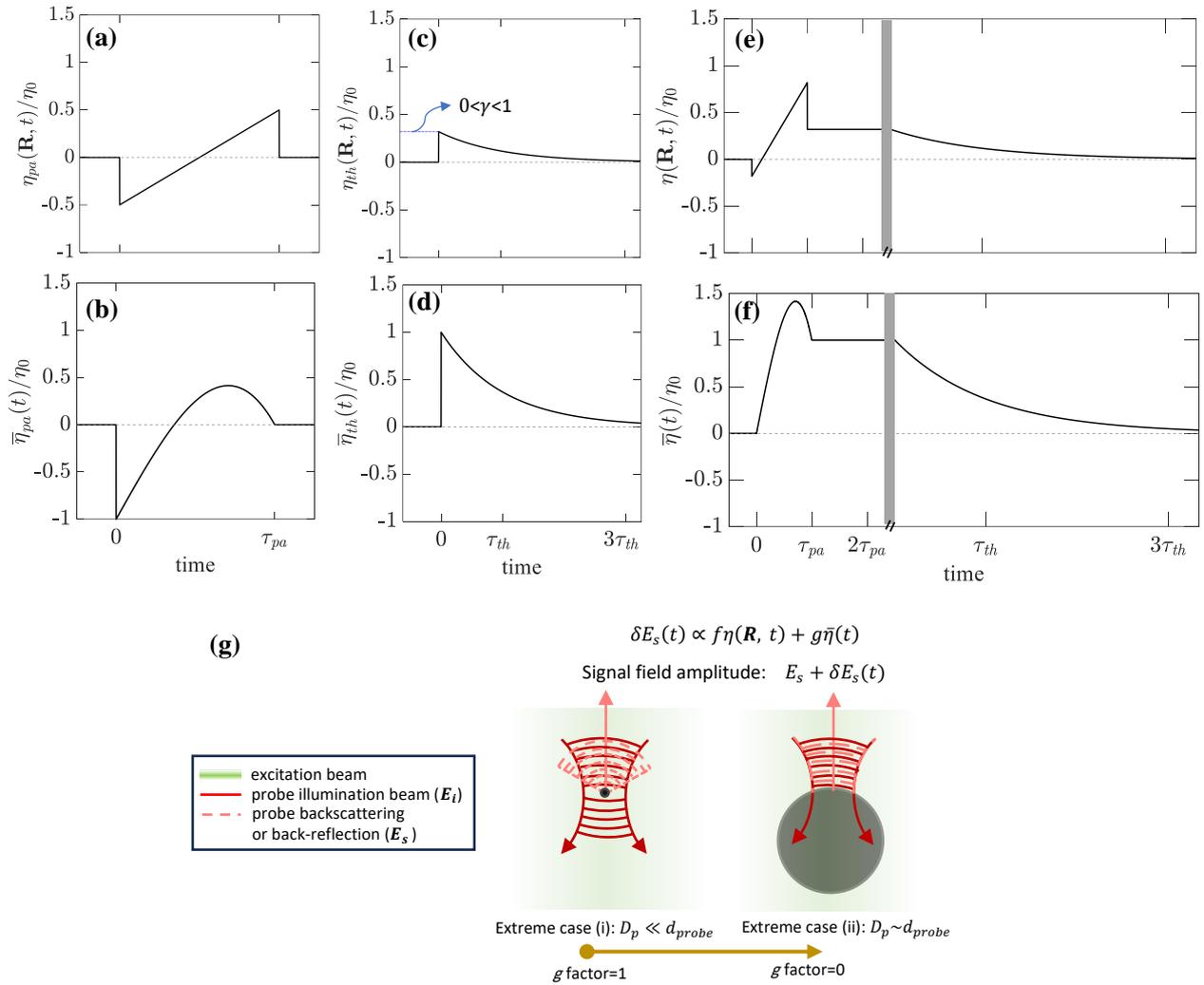

FIG. 3: The relevant strain terms and their relation to the NR-PARS signal. (a) The normalized temporal profile of PA-induced strain at the particle-medium interface, $\eta_{pa}(R,t)$. (b) The normalized temporal profile of PA-induced volume-averaged strain across the object, $\bar{\eta}_{th}(t)$. (c) The normalized temporal profile of thermal-induced strain at the particle medium interface, $\eta_{th}(R,t)$. (d) The normalized temporal profile of thermal-induced volume-averaged strain across the object, $\bar{\eta}_{th}(t)$. (e) The normalized temporal profile of the strain at the particle medium interface, $\eta(R,t)$, superimposed

by strain terms of $\eta_{pa}(R,t)$ and $\eta_{th}(R,t)$. (f) The normalized temporal profile of volume-averaged strain across the object, $\bar{\eta}(t)$, resulting from the sum of the strain terms of $\bar{\eta}_{pa}(t)$ and $\bar{\eta}_{th}(t)$. (g) The final NR-PARS signal is a linear combination of two strain functions, with the contributions of each weighted by the respective *f* factor and *g* factors. While *f* factor can vary between positive and negative values, the g factor is bound between 0 and 1. The upper and lower limit of *f* factor is associated with extreme geometrical case (i) and (ii), respectively.

4. Samples preparation

To verify our concept and model, polystyrene beads embedded in PDMS polymer were used as samples to perform experiments. For sample preparation, the solution of black-dyed polystyrene beads (Polyscience, USA) of diameter 1, 3, 6 and 10 μm was mixed into the PDMS base solution. The samples were stirred and heated overnight to ensure complete evaporation of the water content. Subsequently, the PDMS curing agent, constituting one-tenth of the base solution, was introduced to finalize the preparation of our samples.

5. Experimental setup

The optical system, which is shown in Fig. 1(a), was constructed using the following components. In its excitation submodule, a frequency-doubled laser (IPG photonics YLPP-25-3-50-R) beam with a 5 kHz repetition rate and 2 ps duration was focused by 20×, 0.4 NA (numerical aperture) onto the sample with a focused spot size of ~20 μm (FWHM). In the detection submodule, A 1310 nm fiber-based continuous laser (Thorlabs- SLD1310) was collinearly combined with the excitation laser and co-focused onto the sample via the same objective (focal spot size ~2 μm). The NR-PARS signal (signal field's amplitude modulations) was epi-detected by a photodiode (Thorlabs, PDM480AC-C; bandwidth:1.6 GHz). The output of the photodetector was directed to a fast digitizer with a sampling rate of 20 GSa/s (Keysight MSOX6004A). To improve the signal-to-noise ratio (SNR), 1000 traces were averaged in a single measurement. Sample scanning (Newport Corporation) and signal acquisition were synchronized via data acquisition with a home-built PYTHON code.

6. Results and discussions

6.1. Model demonstrations for mechanical sensing of single micro-spheres

To simulate the theoretical NR-PARS signal, we employed our descriptive model for two absorber-medium combinations: (i) polystyrene sphere (absorbing) embedded in a PDMS medium (non-absorbing), corresponding to our experimental phantom study's sample; (ii) lipid sphere (absorbing) suspended in an aquatic medium (e.g., water/agarose gel) (non-absorbing), representing lipid contrast imaging in cells/tissue. For both absorber-medium combinations, spheres with diameters ($D_p$) of 300 nm and 6 μm are considered, to exemplify the extreme geometrical cases of (i) $D_p \ll d_{probe}$ and (ii) $D_p \sim d_{probe}$, respectively.

The theoretical NR-PARS signal ($\delta E_s(t)$) for polystyrene particles with $D_p \ll d_{probe}$ embedded in PDMS medium is shown in Fig. 4 (a). The signal is dominated by the absorber's volume modulation, that is $\bar{\eta}(t)$. This is because the determined *f* factor (~-0.3) and amplitude of $\eta(R,t)$, respectively, are much smaller compared to *g* factor ((~1)) and amplitude of the strain term $\bar{\eta}(t)$, respectively. For $D_p = 300$ nm, the simulation yields $\tau_{pa} \approx 0.13$ ns and $\tau_{th} \approx 86$ ns. On the other hand, refractive index modulation, dictated by $\eta(\mathbf{R}, t)$), is the sole contributor in NR-PARS signal when $D_p \sim d_{probe}$ [g factor=0 in Eq. (7a)]. As depicted in Fig. 4(b), the signal exhibits a dominant negative sign, indicating diminished reflectivity as non-radiative relaxation occurs. This is because the corresponding $f$ factor (~-0.29) is negative. The initial positive signal [Fig. 4(b)] is characterized by $\gamma \approx 0.3 < 0.5$. For 6 µm polystyrene beads, $\tau_{pa}$ and $\tau_{th}$ are approximated as 2.6 ns and 34.5 µs, respectively.

For the lipid sphere in an aquatic medium, theoretical NR-PARS signal ($\delta E_s(t)$), when $D_p \ll d_{probe}$, is illustrated in Fig. 4(c). Unlike the previous absorber-medium configuration, in addition to the object's volume modulation, the effect of boundary strain $\eta(\mathbf{R}, t)$ is pronounced in this signal. This is because of the determined $f$ factor (~1.04), which is of the same order of magnitude as the $g$ factor (~1). The model calculates $\tau_{pa}$ and $\tau_{th}$ approximately as $\tau_{pa} \approx 0.2$ ns and $\tau_{th} \approx 28$ ns, respectively for $D_p = 300$ nm. When $D_p \sim d_{probe}$, the NR-PARS signal for lipid-water complex, in contrast to polystyrene-PDMS absorber-medium configuration, is dominantly positive. This behavior indicates an increased back-reflection of the probe beam from the lipid sphere as a result of non-radiative relaxation. This is because its corresponding $f$ factor (~1.01) is positive. The initial negative part of the signals in Fig. 4(c) and (d) is the indicator of the associated $\gamma \approx 0.3$ less than 0.5. The model approximates $\tau_{pa}$ and $\tau_{th}$ as 4 ns and 11.2 µs respectively for a 6 µm lipid sphere.

As demonstrated by illustrative examples, regardless of the possible values for $f$ and $g$ factors, NR-PARS signal, according to our theoretical description, consists of a transient PA-induced signal lasting for the duration of $\tau_{pa}$, and a pseudo-static thermal-induced component with a decay constant time of $\tau_{th}$. The underlying reason is that both PA-induced strain terms of $\eta_{pa}(\mathbf{R}, t)$ and $\bar{\eta}_{pa}(t)$ share the same duration of $\tau_{pa}$. Also, both thermal-induced strain terms, $\eta_{th}(\mathbf{R}, t)$ and $\bar{\eta}_{th}(t)$, approximately have the same decay constant time of $\tau_{th}$.

Importantly, the duration of the transient signal for spherical absorbers, that is $\tau_{pa}$, unveil the ratio between the sound speed of the absorber and its diameter. Therefore, this allows for the back-calculation of the sound speed of the absorbing particle, which relates to its elastic properties, provided that the size of the absorber is known. Conversely, $\tau_{pa}$ can also provide information about the size of the absorbing object if its sound speed is known. This is of importance as it potentially allows for the sizing of sub-diffraction-limit objects, offering a new dimension in the applications of NR-PARS.

As mentioned, the model assumes the acoustic impedance of the absorber ($Z_p$) and medium ($Z_m$) nearly equal ($Z_m \approx Z_p$). This approximation makes the effect of PA wave back reflection on the NR-PARS signal negligible, a scenario commonly encountered in diverse absorber-medium combinations in biological samples, tissues, and cells (see the acoustic impedance ($Z=\rho v$) of water and lipid as a biological sample example from table S2). However, in certain absorber-medium complexes, a significant acoustic impedance mismatch may exist between the absorber and medium ($Z_m \neq Z_p$). In these scenarios, the boundary reflects a fraction of emerged traveling PA waves back, quantified by the acoustic reflection coefficient $R_{ac}$, where $R_{ac} = \left|\frac{Z_m - Z_p}{Z_m + Z_p}\right|$. An example of this is our experimental phantom, polystyrene particle in PDMS medium, where $R_{ac}$ is approximately 0.4. The PA propagation simulation, which accounts for the acoustic impedance mismatch, is discussed in Supplementary Note 3 and was performed via FEM solver. Temporal profiles of both interface and volume-averaged pressure amplitudes of the polystyrene particle — which dictate strain terms of $\eta_{pa}(\mathbf{R}, t)$ and $\bar{\eta}_{pa}(t)$, respectively — show the same temporal pattern as depicted in Figs. 2(b) and 2(c) over the $0 < t < \tau_{pa}$ interval (see Fig. S5). However, the wavelets repeat themselves at every $\tau_{pa}$ period, giving rise to periodic PA-induced wavelets in the NR-PARS signal whose amplitude decreases over each period according to the fractional relationship of $R_{ac}$. Therefore, even in scenarios where $Z_m \neq Z_p$, the previously mentioned mechanical and geometrical properties of the absorbing object can still be deduced from the time-resolved NR-PARS signal.

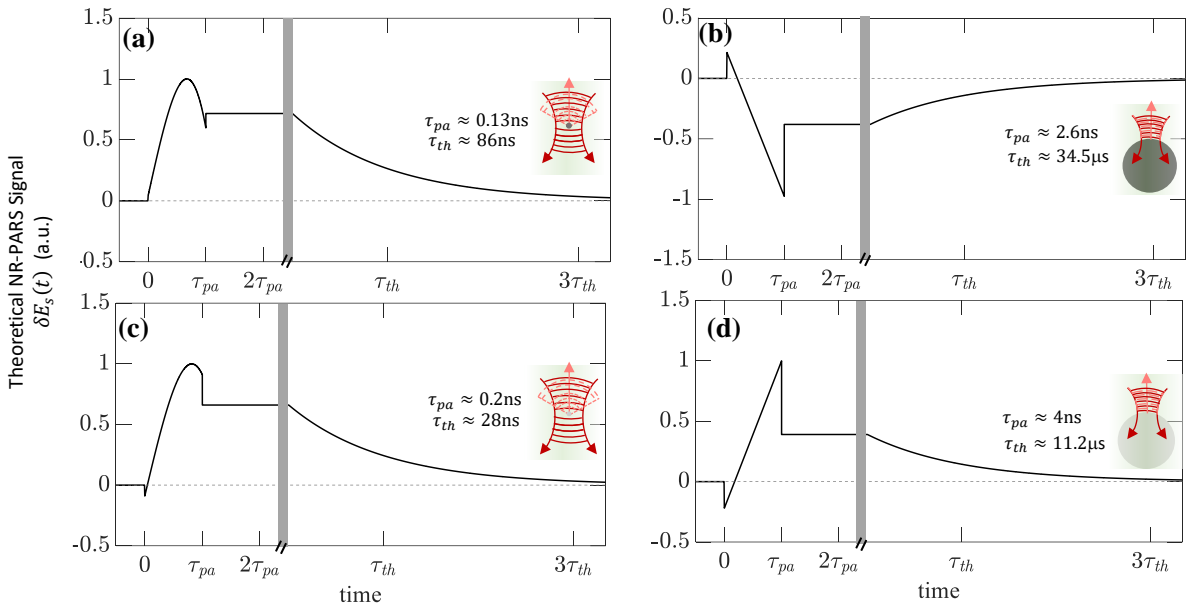

**FIG. 4: The theoretical NR-PARS signals of two distinct absorber-medium complexes at two geometrical regimes, obtained from superposition of the transient PA-induced with duration of $\tau_{pa}$ and thermal-induced signal with decay constant time of $\tau_{th}$. Time-resolved theoretical NR-PARS signal of a (a) 300 nm and (b) 6 μm diameter polystyrene in PDMS medium. Theoretical NR-PARS signals from lipid droplets of (c) $D_p$ =300 nm and (d) $D_p$ =6 μm suspended in aquatic medium. The corresponding geometrical configurations of the absorber and the probe's focus are illustrated in insets for each signal.**

6.2. Experimental verification for spherical polystyrene absorbers

The NR-PARS submodule presented in Fig. 1(a) was built to record the signals from individual polystyrene particles of various sizes suspended in a PDMS medium. These measurements were performed to verify our descriptive model. To locate the center of single beads, a scattering scan was employed, in which the focal plane of the probe beam was positioned 10 μm below the surface of the PDMS block. After identifying the bead's center, the stage position was adjusted to align the bead center with the center of excitation and probe beams. The excitation laser was targeted to illuminate the entire individual polystyrene sphere to resemble the model's assumptions. The excitation fluence was set to 5 mJ/cm$^2$.

Figure 5 (a)-(d) display the experimental NR-PARS signals from polystyrene beads of diameters of 1, 3, 6, and 10 mm, respectively, by gray lines for 30 ns of observation. Alongside, the corresponding band-limited modeled signal, derived from theoretical NR-PARS signal [Eq. (9); assuming *g*=0] after incorporating the photodiode's frequency response, is shown. The signals having a dominant negative sign are inversed for better illustration. Also, in the experimental signals, the periodic wavelets mediated by the PA wave reflections may be obscured by noise and bandwidth-mediated ringing effects. That is due to the weakened nature of the reflected PA pressure, characterized by the reflection coefficient $R_{ac}$ ≈0.4. During the 30 ns of observation, the thermal-induced signal is regarded as static. Overall, a good agreement was observed between the simulation results from the developed model and the experimental results. The extended experimental NR-PARS signal and corresponding model results, with a longer observation time, are shown in Fig. S6.

As discussed previously, in the theoretical NR-PARS signal (infinite bandwidth), $\tau_{pa}$ reveals the ratio of the absorber's diameter to its speed of sound. However, in realistic scenarios involving band-limited signals and noise, the rise time ($\tau_{rt}$) is introduced as an alternative parameter instead of $\tau_{pa}$. Rise time ($\tau_{rt}$) is defined as the duration required for a signal to change from 10% to 90% of its step height (i.e., the difference between its minimum and maximum values). This definition is employed to mitigate the effect of inherent noise and bandwidth-mediated distortion in the signal

Figure 6 shows the rise time for experimental as well as band-limited modeled signals for different sizes of polystyrene beads. A deviation is observed between the rise times (experimental and band-limited modeled) and $\tau_{pa}$ because of the limited bandwidth of the photodetector and noise. However, a linear correlation exists between $\tau_{pa}$ and $\tau_{rt}$ which can be mapped according to the relation $\tau_{rt} = m\tau_{pa} + c$, where $m$ and $c$ are scaling and offset parameters respectively, related to detection set-up only.

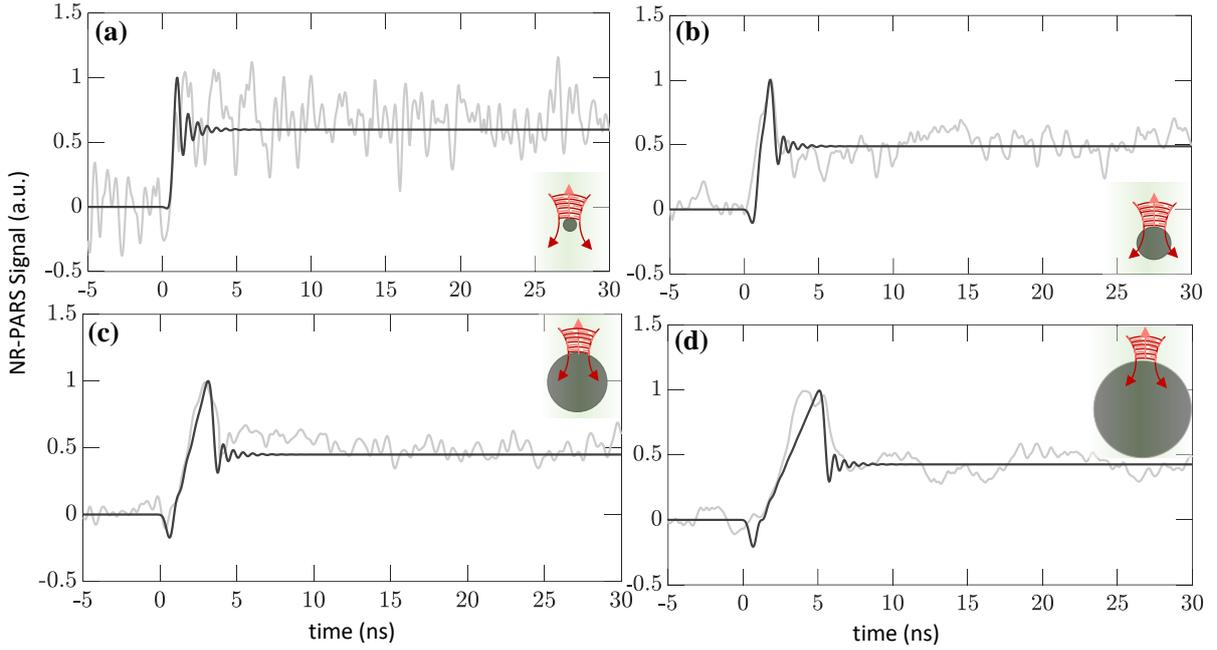

FIG. 5: The experimental NR-PARS signals (grey line) and bandlimited modeled signals (black line) for polystyrene beads with sizes (a) 1 μm (b) 3 μm (c) 6 μm and (d) 10 μm. The corresponding geometrical configurations of the absorber and the probe's focus are illustrated in insets for each signal.

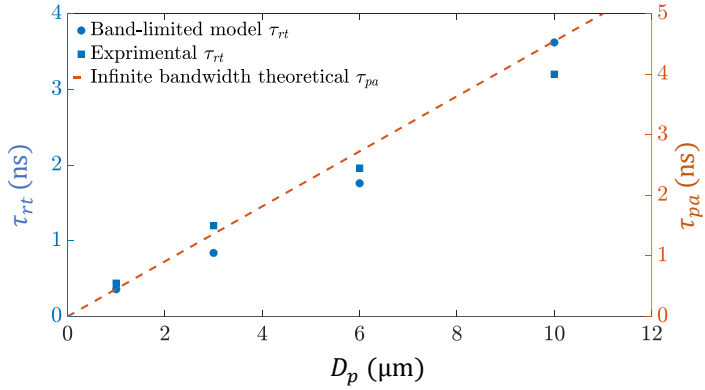

FIG. 6: Linear relationship between experimental and modeled rise times versus theoretical relaxation time for various polystyrene bead sizes.

### 7. Current Limitations and Future Path

In the current setup, the bandwidth of the detection is in the range of 30 kHz to 1.6 GHz, imposing a limitation in capturing PA temporal profiles for micro-objects smaller than 1 μm. To extend the applicability of this methodology to mechanical imaging of submicron absorbers, Optical gating via ultrashort pulses and a delay line can be used instead of a CW probe laser. This enables terahertz detection with an observation time of $O(1)$ ns.

Additionally, in the present CW probe setup, expanding the photodetector's bandwidth to tens of GHz could ensure the accurate and distortion-free capture of PA profiles for submicron absorbers. However, incorporating higher bandwidth into the setup poses a challenge, that is increased noise levels as noise $\propto \sqrt{BW}$, where $BW$ is the bandwidth of the photodetector. To mitigate this challenge and improve SNR, three methods can be considered. First, surpassing the intensity noise can be achieved by using CW probe lasers with lower intensity noise and/or adopting balanced detection. Secondly, high-intensity pump lasers can be employed and consequently, the backscattered light intensity will be higher to overcome the background noise of the detector, i.e., dark current noise. Lastly, signal detection from forward scattering in transmission modes can be helpful as forward scattering is significantly larger than backward scattering in micron-scale scatterers, according to Mie theory. The current setup requires multiple averaging to improve the SNR mainly due to operating in the dark noise regime of the photodetector. However, achieving shot-noise-limited detection would reduce reliance on averaging procedures, enabling the possibility of single-shot signal detection. With these improvements, an experimentally extracted PA temporal profile could more closely resemble theoretical signals, significantly enhancing the accuracy and resolution of NR-PARS-based mechanical imaging.

## 8. Conclusions

In conclusion, we have presented a proof of concept for mechanical sensing through the characterization of NR-PARS signal arising from non-radiative relaxation following ultrashort pulsed laser absorption. A descriptive mechanism model has been presented for NR-PARS signal which allows for the extraction of minimally distorted photoacoustic temporal profile emitted from individual absorbing micro/nano-objects. Not only that but also the presented model can be used to optimize the parameters of the absorber and medium as well as the optical setup configuration to improve detection efficiency of PARS, an absorption-based microscopy. Experiments on individual polystyrene microbeads of various sizes embedded in PDMS block show good agreement with our descriptive model. The NR-PARS-based mechanical assessment exhibits significant potential for biomechanical imaging, offering advantages over existing techniques such as Brillouin scattering (spontaneous/stimulated), given its larger cross-section and, consequently, higher SNR signal. Additionally, we have discussed current limitations in our experiment and proposed improvements. By implementing future improvements in NR-PARS submodule, mechanical imaging of micron and submicron absorbers will become possible with much higher SNR. Furthermore, our method, when combined with prior knowledge of the absorber's speed of sound and shape, could potentially enable the measurement of sub-diffraction-limit object sizes.

**References**

[1] A. N. Ketene, E. M. Schmelz, P. C. Roberts, and M. Agah, "The effects of cancer progression on the viscoelasticity of ovarian cell cytoskeleton structures," *Nanomedicine: Nanotechnology, Biology and Medicine,* vol. 8, no. 1, pp. 93-102, 2012.

[2] C. F. Guimarães, L. Gasperini, A. P. Marques, and R. L. Reis, "The stiffness of living tissues and its implications for tissue engineering," *Nature Reviews Materials,* vol. 5, no. 5, pp. 351-370, 2020.

[3] E. Gil-Santos *et al.*, "Optomechanical detection of vibration modes of a single bacterium," *Nature nanotechnology,* vol. 15, no. 6, pp. 469-474, 2020.

[4] K. Li and Y. N. Jan, "Experimental tools and emerging principles of organellar mechanotransduction," *Trends in Cell Biology,* 2025.

[5] S.-J. Tang *et al.*, "Single-particle photoacoustic vibrational spectroscopy using optical microresonators," *Nature Photonics,* vol. 17, no. 11, pp. 951-956, 2023.

[6] S. Mattana *et al.*, "Non-contact mechanical and chemical analysis of single living cells by microspectroscopic techniques," *Light: Science & Applications,* vol. 7, no. 2, pp. 17139-17139, 2018.

[7] F. Palombo *et al.*, "Chemico-mechanical imaging of Barrett's oesophagus, J. Biophotonics 9 (2016) 694–700," ed.

[8] L. Qiu *et al.*, "A high-precision multi-dimensional microspectroscopic technique for morphological and properties analysis of cancer cell," *Light: Science & Applications,* vol. 12, no. 1, p. 129, 2023.

[9] X. Gao, X. Li, and W. Min, "Absolute stimulated Raman cross sections of molecules," *The journal of physical chemistry letters,* vol. 14, no. 24, pp. 5701-5708, 2023.

[10] F. Yang *et al.*, "Pulsed stimulated Brillouin microscopy enables high-sensitivity mechanical imaging of live and fragile biological specimens," *Nature Methods,* vol. 20, no. 12, pp. 1971-1979, 2023.

[11] M. Nikolić and G. Scarcelli, "Long-term Brillouin imaging of live cells with reduced absorption-mediated damage at 660nm wavelength," *Biomedical optics express,* vol. 10, no. 4, pp. 1567-1580, 2019.

[12] Y. Bai *et al.*, "Ultrafast chemical imaging by widefield photothermal sensing of infrared absorption," *Science advances,* vol. 5, no. 7, p. eaav7127, 2019.

[13] K. Huang and A. Rhys, "Theory of light absorption and non-radiative transitions in F-centres," *Proceedings of the Royal Society of London. Series A. Mathematical and Physical Sciences,* vol. 204, no. 1078, pp. 406-423, 1950.

[14] J. Yin *et al.*, "Nanosecond-resolution photothermal dynamic imaging via MHZ digitization and match filtering," *Nature Communications,* vol. 12, no. 1, p. 7097, 2021.

[15] Y. Liu, C. Tao, and X. Liu, "Photoacoustic multispectral elastography based on the photoacoustic oscillation effect for microelastomers in deep tissue," *Physical Review Applied,* vol. 20, no. 2, p. 024035, 2023.

[16] F. Yang, Z. Chen, P. Wang, and Y. Shi, "Phase-domain photoacoustic mechanical imaging for quantitative elastography and viscography," *IEEE Transactions on Biomedical Engineering,* 2024.

[17] Y. Yuan *et al.*, "Photoacoustic remote sensing elastography," *Optics Letters,* vol. 48, no. 9, pp. 2321-2324, 2023.

[18] M. S. Singh and A. Thomas, "Photoacoustic elastography imaging: a review," *Journal of biomedical optics,* vol. 24, no. 4, pp. 040902-040902, 2019.

[19] M. J. Moore *et al.*, "Photoacoustic F-mode imaging for scale specific contrast in biological systems," *Communications Physics,* vol. 2, no. 1, p. 30, 2019.

[20] H. Li, B. Dong, Z. Zhang, H. F. Zhang, and C. Sun, "A transparent broadband ultrasonic detector based on an optical micro-ring resonator for photoacoustic microscopy," *Scientific reports,* vol. 4, no. 1, p. 4496, 2014.

[21] A. A. Plumb, N. T. Huynh, J. Guggenheim, E. Zhang, and P. Beard, "Rapid volumetric photoacoustic tomographic imaging with a Fabry-Perot ultrasound sensor depicts peripheral arteries and microvascular vasomotor responses to thermal stimuli," *European radiology,* vol. 28, pp. 1037-1045, 2018.

[22] Y. Shan, Y. Dong, W. Song, and X. Yuan, "Spectroscopically resolved photoacoustic microscopy using a broadband surface plasmon resonance sensor," *Applied Physics Letters,* vol. 120, no. 12, 2022.

[23] E. Hysi, M. J. Moore, E. M. Strohm, and M. C. Kolios, "A tutorial in photoacoustic microscopy and tomography signal processing methods," *Journal of Applied Physics,* vol. 129, no. 14, 2021.

[24] D. Zhang, C. Li, C. Zhang, M. N. Slipchenko, G. Eakins, and J.-X. Cheng, "Depth-resolved mid-infrared photothermal imaging of living cells and organisms with submicrometer spatial resolution," *Science advances,* vol. 2, no. 9, p. e1600521, 2016.

[25] D. Zhang *et al.*, "Bond-selective transient phase imaging via sensing of the infrared photothermal effect," *Light: Science & Applications,* vol. 8, no. 1, p. 116, 2019.

[26] S. Berciaud, L. Cognet, G. A. Blab, and B. Lounis, "Photothermal Heterodyne Imaging of Individual Nonfluorescent Nanoclusters<? format?> and Nanocrystals," *Physical review letters,* vol. 93, no. 25, p. 257402, 2004.

[27] P. Hajireza, W. Shi, K. Bell, R. J. Paproski, and R. J. Zemp, "Non-interferometric photoacoustic remote sensing microscopy," *Light: Science & Applications,* vol. 6, no. 6, pp. e16278-e16278, 2017.

[28] T. Simmons, S. J. Werezak, B. R. Ecclestone, J. E. Tweel, H. Gaouda, and P. H. Reza, "Label-Free Non-Contact Vascular Imaging using Photon Absorption Remote Sensing," *IEEE Transactions on Biomedical Engineering,* 2024.

[29] M. Boktor, J. E. Tweel, B. R. Ecclestone, J. A. Ye, P. Fieguth, and P. Haji Reza, "Multi-channel feature extraction for virtual histological staining of photon absorption remote sensing images," *Scientific Reports,* vol. 14, no. 1, p. 2009, 2024.

[30] B. R. Ecclestone, K. Bell, S. Sparkes, D. Dinakaran, J. R. Mackey, and P. Haji Reza, "Label-free complete absorption microscopy using second generation photoacoustic remote sensing," *Scientific Reports,* vol. 12, no. 1, p. 8464, 2022.

[31] J. E. Tweel *et al.*, "Photon absorption remote sensing imaging of breast needle core biopsies is diagnostically equivalent to gold standard H&E histologic assessment," *Current Oncology,* vol. 30, no. 11, pp. 9760-9771, 2023.

[32] J. E. Tweel, B. R. Ecclestone, M. Boktor, D. Dinakaran, J. R. Mackey, and P. H. Reza, "Automated whole slide imaging for label-free histology using photon absorption remote sensing microscopy," *IEEE Transactions on Biomedical Engineering,* vol. 71, no. 6, pp. 1901-1912, 2024.

[33] A. Prost, F. Poisson, and E. Bossy, "Photoacoustic generation by a gold nanosphere: From linear to nonlinear thermoelastics in the long-pulse illumination regime," *Physical Review B,* vol. 92, no. 11, p. 115450, 2015.

[34] L. V. Wang and H.-i. Wu, *Biomedical optics: principles and imaging*. John Wiley & Sons, 2007.

[35] A. P. Sarvazyan, O. V. Rudenko, S. D. Swanson, J. B. Fowlkes, and S. Y. Emelianov, "Shear wave elasticity imaging: a new ultrasonic technology of medical diagnostics," *Ultrasound in medicine & biology,* vol. 24, no. 9, pp. 1419-1435, 1998.

[36] S. E. Bialkowski, N. G. Astrath, and M. A. Proskurnin, *Photothermal spectroscopy methods*. John Wiley & Sons, 2019.

[37] D. F. Hays and H. N. Curd, "Heat conduction in solids: Temperature-dependent thermal conductivity," *International Journal of Heat and Mass Transfer,* vol. 11, no. 2, pp. 285-295, 1968.

[38] W.-C. Tsai and R. J. Pogorzelski, "Eigenfunction solution of the scattering of beam radiation fields by spherical objects," *Journal of the Optical Society of America,* vol. 65, no. 12, pp. 1457-1463, 1975.

[39] C. F. Bohren and D. R. Huffman, *Absorption and scattering of light by small particles*. John Wiley & Sons, 2008.